# Generating and Detecting Various Types of Fake Image and Audio Content: A Review of Modern Deep Learning Technologies and Tools

Arash Dehghani[1*], Hossein Saberi[2]

[1,2] Faculty of Artificial Intelligence and Cognitive Science, Imam Hossein Comprehensive University
[1] a.dehghani@ihu.ac.ir, [2] hsaberi @ihu.ac.ir

## Abstract

This paper reviews the state-of-the-art in deepfake generation and detection, focusing on modern deep learning technologies and tools based on the latest scientific advancements. The rise of deepfakes, leveraging techniques like Variational Autoencoders (VAEs), Generative Adversarial Networks (GANs), Diffusion models and other generative models, presents significant threats to privacy, security, and democracy. This fake media can deceive individuals, discredit real people and organizations, facilitate blackmail, and even threaten the integrity of legal, political, and social systems. Therefore, finding appropriate solutions to counter the potential threats posed by this technology is essential. We explore various deepfake methods, including face swapping, voice conversion, reenactment and lip synchronization, highlighting their applications in both benign and malicious contexts. The review critically examines the ongoing "arms race" between deepfake generation and detection, analyzing the challenges in identifying manipulated contents. By examining current methods and highlighting future research directions, this paper contributes to a crucial understanding of this rapidly evolving field and the urgent need for robust detection strategies to counter the misuse of this powerful technology. While focusing primarily on audio, image, and video domains, this study allows the reader to easily grasp the latest advancements in deepfake generation and detection.

# 1. Introduction

Deep learning has successfully been applied to solve various complex problems, ranging from big data analysis to computer vision and human-level control. However, advancements in deep learning have also been used to create software that can threaten privacy, democracy, and national security. One such deep learning-based application that has recently emerged is the deepfake, first appearing in 2017 [1]. Deepfake algorithms, lexically derived from "deep" learning and "fake," can create fake images, videos, audio, and text that are difficult for humans to distinguish from authentic samples [2]. Deepfakes typically refer to the manipulation of existing media or the generation of new (synthetic) media using machine learning-based approaches. The most commonly discussed deepfake data are fake facial images, fake speech synthesis, and fake videos that incorporate both fake images and speech. Voice conversion (VC) is a technology that can be used to alter an audio to sound as if it were spoken by a different person (target) than the original speaker (source). Previously, other methods such as traditional visual effects or computer graphics approaches were used to generate fake content. However, recently, the common underlying mechanism for creating deepfakes is deep learning models like Variational Autoencoders (VAEs), Generative Adversarial Networks (GANs), and Diffusion models, widely used in computer vision. In general, the generation of fake content has shifted from traditional graphics-based methods to deep learning-based approaches [3]. With advancements in deep learning, techniques primarily provided by autoencoders and generative adversarial networks have achieved remarkable generative results. Recently, the emergence of diffusion models with powerful generation capabilities has spurred a new wave of research. In a narrow sense, deepfakes refer to facial forgery. Face swapping is a remarkable task in generating face-based fake content and is accomplished by transferring the source face to the destination while preserving the facial movements and expressions of the destination [4]. In a broader sense, it also includes the entire image composition and is not limited to the face, widely termed AIGC (AI-generated content) [5].

* Corresponding author, E-mail address: a.dehghani@ihu.ac.ir (Arash Dehghani)

Since creating realistic digital humans has positive consequences, positive uses of deepfakes exist, such as their applications in visual effects, digital avatars, Snapchat filters, creating voices for people who have lost their voices or updating parts of films without reshooting, video games, virtual reality, film productions, and entertainment, realistic dubbing of foreign films, education through the revival of historical figures, trying on clothes while shopping, etc. [2]. However, when used maliciously, deepfakes can have detrimental consequences for political and social organizations, including decreased public trust in institutions, political manipulation, influencing elections, destabilizing public trust in political institutions, damaging the reputations of prominent individuals, and influencing public opinion [6]. Furthermore, this technology allows malicious actors to create and distribute non-consensual explicit content to cause harassment and reputational damage or create convincing forgeries of individuals to deceive others for financial or personal gain. Moreover, the increased use of deepfakes poses a serious issue in digital forensics, contributing to a general crisis of trust and authenticity in digital evidence used in legal proceedings, and can further lead to malicious uses such as blackmail, extortion, forgery, identity theft, character assassination, and deepfake pornography [7 ,1]. Especially since creating manipulated images and videos is now much easier, as only a photograph of the target's identity or a short video of an individual is needed to generate fake content [8].

The negative aspect of this new technique highlights another popular area of study: the detection of deepfake-generated content, one example of which aims to identify fake faces from real ones. With the rapid development of deepfake-related studies in the community, both sides (generation and detection) have formed a neck-and-neck competition, pushing each other's advancements and inspiring new avenues. Besides deepfake generation, detection technologies are constantly evolving to mitigate the potential misuse of deepfakes, such as privacy violations and phishing attacks [3]. Cybersecurity in the deepfake domain requires not only research into detection but also investigation into generation methods [4]. Therefore, finding truth in the digital realm has become increasingly crucial. This is even more challenging when dealing with deepfakes because they are primarily used for malicious purposes, and almost anyone can create fake content these days using readily available deepfake tools. Numerous methods for deepfake detection have been proposed to date [2]. The importance of this work is such that the U.S. Defense Advanced Research Projects Agency (DARPA) has initiated a research program in the area of media forensics (called MediFor) to accelerate the development of methods for detecting fake digital visual media.

This paper comprehensively reviews the latest advancements in deepfake generation and detection, offering a summary analysis of the current state-of-the-art in this rapidly evolving field. The next section outlines the motivation for this work. Chapter three will delve into the history of the technology, roadmap, and related work, outlining the tools and methods used in deepfake generation and detection. Chapter four will describe the various types of deepfakes focusing on four well-known deepfake research areas: face swapping, voice conversion, lip synchronization, talking face generation, and facial feature editing, and chapter five will discuss deepfake detection. Finally, a summary and conclusion will be presented in the last chapter.

## 2. Motivation

Research in deepfake generation and detection is driven by several crucial motivations, reflecting both technological advancements and societal implications.

Recently, the proliferation of artificially generated content has made deepfake generation and detection techniques a compelling research area. The increasing accessibility of content generation tools has made them a convenient option for illicit activities worldwide. While some positive uses exist, they are predominantly used to create and disseminate fake content. As malicious and harmful applications proliferate faster than beneficial ones, the study of deepfakes has become critically important in the current climate [9]. This research is vital not only for academic purposes but also for practical applications in law enforcement and digital content verification. Furthermore, understanding the motivations behind deepfake creation, ranging from entertainment to malicious intent, can inform strategies for regulation and public awareness, ultimately fostering a more informed society capable of navigating the complexities introduced by this technology. This work aims to provide a comprehensive discussion of the fundamental principles of deepfakes in the realms of audio, video, and image generation, as well as current tools for deepfake detection.

## 3. Background and related work

This section begins with a brief introduction to the background and related works. We will then describe the tools used for generating deepfake content, considering both academic works and widely used open-source software. The manipulation of image and video content, developed in the 19th century and soon applied to moving images, is not new. For this purpose, several dedicated software tools such as Adobe Photoshop and Adobe Lightroom have been available for decades [1]. Deepfake technology has been developed by researchers in academic institutions since the 1990s and later by amateurs in online communities. Recently, these methods have been adopted by industry. Before deepfakes, images or videos were manipulated using image/video splicing, also known as copy-move forgery [7]. For images, specific parts are cut and pasted onto another area. Thus, images are manipulated by overwriting another image. However, these methods required significant expertise, were time-consuming, and often resulted in noticeable artifacts. The advent of deep learning, specifically Generative Adversarial Networks (GANs), Variational Autoencoders (VAEs), and more recently, diffusion models, revolutionized the creation of realistic fake content. These techniques, initially developed within academic research, have been increasingly adopted by both amateur and professional communities, significantly lowering the barrier to entry for creating convincing deepfakes [10].

The accessibility of deepfake creation has been further enhanced by the proliferation of open-source tools and readily available datasets. Software packages like Roop-unleashed, LivePortrait, DeepFaceLab[4], FakeApp, etc. provide user-friendly interfaces and pre-trained models, enabling individuals with limited technical skills to generate realistic deepfakes. Tables 1 through 3 fully detail the deepfake generation tools examined in this paper. Publicly accessible datasets, such as CelebA-HQ, FFHQ, VoxCeleb, etc., provide large quantities of high-quality data for training and fine-tuning these models. This ease of access has fueled the rapid spread of deepfakes across various platforms and applications. Early deepfake methods largely focused on facial manipulation, particularly face swapping, which involves transferring a source face onto a target face while preserving the target's expressions and movements. However, the scope of deepfake technology has expanded significantly. Audio deepfakes, achieved through techniques such as Seed-VC, KNN-VC, HierSpeechpp, WaveNet, Tacotron, etc., pose a significant threat. These methods can generate highly realistic synthetic speech, enabling impersonation, voice cloning, and the creation of convincing audio recordings that never actually occurred. The challenges in detecting audio deepfakes are often different from those in visual deepfakes, requiring specialized techniques that analyze acoustic features and subtle nuances in speech patterns. Furthermore, deepfake generation is evolving beyond images and audio. Researchers are exploring the creation of synthetic videos with realistic body movements and interactions, expanding into other modalities like text and even video games. The rapid pace of innovation in deepfake generation necessitates a constant evolution in detection techniques, and this ongoing "arms race" between generation and detection is driving advancements in both fields. This review aims to comprehensively examine these diverse approaches and challenges, considering both the technical underpinnings and the broader societal impact of this rapidly evolving technology.

# 4. Material and Methods

Deepfakes primarily utilize deep neural networks to manipulate video, image, audio, and text content [7]. Deep learning techniques such as autoencoders, GANs, and CNNs play a crucial role in deepfake creation, enabling the generation of highly realistic synthetic media. Understanding these technologies is vital for developing effective detection strategies to combat associated risks. In the following section, we will introduce some important deep neural network architectures.

## 4.1. Artificial Neural Networks

Artificial Neural Networks (ANNs) are computational algorithms designed to mimic the function of the human brain. Inspired by the structure and function of neurons, these networks are primarily developed for tasks involving learning from data. ANNs consist of three main layers: an input layer, hidden layers, and an output layer [11]. The input layer feeds data into the system, while the hidden layers process and extract features from the data. The output layer provides predictions or classifications. The operation of ANNs is divided into two stages: feedforward and backpropagation. In the feedforward stage, data is passed from input to output, and in the backpropagation stage, the weights of the connections between neurons are updated to minimize error [12]. Neural networks are widely used in applications such as image recognition, natural language processing, and financial prediction, and due to their high capacity for learning from complex data, they have become one of the most important tools in machine learning.

## 4.2. Deep learning

Deep learning is a subfield of machine learning that focuses on algorithms inspired by the structure and function of the brain, particularly artificial neural networks. In recent years, it has gained significant traction due to its remarkable ability to process large amounts of data and perform complex tasks across various domains. Deep learning uses neural networks with multiple layers (hence the term "deep") to model complex patterns in data [13]. Unlike traditional machine learning methods that often require manual feature extraction, deep learning automatically discovers representations from raw data and is particularly effective for tasks such as image and speech recognition, natural language processing, and more.

This scientific field has seen rapid advancements, especially in generative models capable of producing high-quality synthetic data. For instance, deep learning models can generate realistic images from text descriptions or produce human-like speech. Advances in deep learning have not only enhanced existing technologies but have also opened up new avenues for research and application across various scientific disciplines. Some common architectures of deep learning networks include fully connected networks, deep belief networks, recurrent neural networks, convolutional neural networks, generative adversarial networks, transformers, and neural radiance fields. These architectures have been applied in fields including computer vision, speech recognition, natural language processing, machine translation, bioinformatics, drug design, medical image analysis, climate science, materials inspection, and board game applications, where they have produced results comparable to and, in some cases, surpassing the performance of human experts [14].

## 4.3. Convolutional Neural Networks

Convolutional Neural Networks (CNNs) are a specialized type of deep learning architecture primarily used for analyzing visual data. They have become a cornerstone of modern computer vision, enabling applications such as image classification, object detection, and segmentation. A CNN is a feedforward neural network that processes data using a grid-like topology, typically applied to images. CNNs consist of multiple layers, including convolutional layers, pooling layers, and fully connected layers. Each layer plays a crucial role in feature extraction and transformation, allowing the network to learn hierarchical representations of the input data. CNNs utilize the backpropagation algorithm to optimize filter weights during training, enabling them to effectively learn from labeled datasets. This capability of automatic feature extraction distinguishes CNNs from traditional machine learning methods, which often require manual feature engineering. The architecture is designed to handle variations in input data through mechanisms such as weight sharing and local connectivity, enhancing translational invariance—meaning features are detectable regardless of their position within the image. The concept of CNNs was introduced by Yann LeCun in 1988 with his LeNet architecture, designed for character recognition tasks. Since then, CNNs have evolved significantly and are now foundational in deep learning applications across various domains [15]. However, the lineage of this neural network type can be traced back to the architecture introduced by Kunihiko Fukushima in 1979, which also introduced max-pooling, a now popular down sampling method [16]. The Time Delay Neural Network (TDNN), introduced by Alex Waibel in 1987, applied CNNs to phoneme recognition, utilizing convolution, weight sharing, and backpropagation [17]. In summary, convolutional neural networks represent a powerful approach to understanding visual data through automatic feature extraction and hierarchical learning.

## 4.4. Recurrent Neural Networks

Recurrent Neural Networks (RNNs) are a specialized class of artificial neural networks designed for processing sequential data, making them particularly effective for tasks where contextual and temporal dependencies are crucial. One origin of RNNs lies in statistical mechanics. In 1972, Shunichi Amari proposed a statistical method to modify the Ising model's weights via Hebbian learning as a model of associative memory, adding a learning component. This work was popularized as the Hopfield network by John Hopfield (1982) [18]. Further developed in the 1980s by researchers such as David Rumelhart, Geoffrey Hinton, and Ronald J. Williams, RNNs have significantly impacted fields like Natural Language Processing (NLP), speech recognition, and time series analysis. Unlike traditional feedforward neural networks that process inputs and outputs independently, RNNs maintain a hidden state that acts as a memory of previous inputs. This hidden state is updated at each time step based on the current input and the previous hidden state, allowing RNNs to capture dependencies over time. This feedback loop enables RNNs to make predictions dependent on prior context, making them suitable for tasks such as predicting the next word in a sentence or analyzing time series data.

The architecture of RNNs comprises interconnected neurons organized into input, output, and hidden layers. The hidden layer processes sequential data step-by-step while retaining information from previous steps through its recurrent connections. This design allows RNNs to efficiently handle variable-length input sequences. RNNs are trained using a method called Backpropagation Through Time (BPTT), which extends the traditional backpropagation algorithm to account for the temporal nature of the data. BPTT computes gradients at each time step and adjusts network weights accordingly, facilitating learning in sequences. RNNs can be categorized into various types based on their architecture and application:

- Simple RNNs: The simplest form of RNN, suffering from limitations such as the vanishing gradient problem.
- Long Short-Term Memory (LSTM): A more advanced variant designed to overcome the vanishing gradient problem by introducing memory cells capable of storing information over longer periods.
- Gated Recurrent Units (GRUs): A simplified version of LSTMs that combines the forget and input gates into a single update gate [19].

- Bidirectional RNNs: These networks process data in both forward and backward directions, improving contextual understanding by considering future inputs alongside past inputs.

In summary, RNNs represent a significant advancement in neural network architectures, particularly for sequential data processing. Their ability to retain information over time makes them essential for many modern AI applications, although they are increasingly being complemented or replaced by transformer-based models in specific domains due to efficiency considerations.

## 4.5. Autoencoders

Autoencoders are a type of artificial neural network designed to learn efficient representations of input data without requiring labeled outputs. The architecture typically consists of three main components: an encoder, a bottleneck (or latent space), and a decoder. The encoder reduces the dimensionality of the input data by mapping it to a lower-dimensional latent space representation, capturing essential features while discarding less relevant information. The bottleneck is a layer containing the compressed representation of the input data; this layer acts as a constraint, forcing the model to learn a more compact representation. The decoder attempts to reconstruct the original input from the latent representation, aiming to minimize the reconstruction error, often measured using loss functions such as Mean Squared Error (MSE) or binary cross-entropy.

Several variations of autoencoders exist, each aiming to leverage learned features for desirable properties. Examples include regularized autoencoders (sparse, denoising, and contractive autoencoders), which are effective in learning representations for subsequent classification tasks, and variational autoencoders, which can be used as generative models. Autoencoders find applications in numerous problems, including face recognition, feature extraction, anomaly detection, and word embedding learning. In terms of data synthesis, autoencoders can also be used to generate new data samples similar to the (training) input data.

## 4.6. Variational Autoencoders

Variational Autoencoders (VAEs) are a class of deep generative models that have garnered significant attention in the machine learning field due to their ability to learn complex data distributions and generate novel data samples. Similar to standard autoencoders, VAEs aim to learn a compressed representation of the input data in a latent space. However, VAEs impose a prior distribution (often a simple Gaussian) on the latent space, encouraging nearby points in the latent space to generate similar data points in the input space. This allows the encoder to generate multiple new data points from a single input, all sampled from the learned distribution. VAEs are designed to learn a probabilistic representation of the input data through the latent space. The primary goal is to encode high-dimensional input data into a lower-dimensional latent space while preserving the essential characteristics of the original data. This is achieved through two main components: an encoder and a decoder. The encoder maps the input data to a distribution in the latent space, typically assumed to be Gaussian. The output consists of parameters (mean and variance) defining this distribution. The decoder reconstructs the original data from samples drawn from the latent representation, effectively generating new data points.

Training Variational Autoencoders involves maximizing a variational lower bound on the log-likelihood of the observed data. This is achieved using variational inference techniques, which allow for efficient approximation of the posterior distribution of the latent variables given the observed data. The loss function consists of two terms:

- Reconstruction Loss: Measures how well the decoder can reconstruct the original input from the latent representation.
- KL Divergence Loss: Regularizes the learned distribution to be close to the prior distribution (typically a standard normal distribution), ensuring smoothness of the latent space.

## 4.7. Transformers

The Transformer architecture is a groundbreaking deep learning model introduced in the seminal paper "Attention is All You Need" by Vaswani et al. in 2017. It revolutionized the field of Natural Language Processing (NLP) and has since been adapted for various applications, including image processing and reinforcement learning. The Transformer model is fundamentally based on an encoder-decoder structure, allowing it to efficiently perform sequence-to-sequence tasks. Unlike traditional models relying on Recurrent Neural Networks (RNNs) or Convolutional Neural Networks (CNNs), Transformers utilize a mechanism called self-attention for parallel processing of input data, significantly improving computational efficiency and performance on long-range dependencies. Input text is first tokenized into smaller units, which are then converted into numerical vectors via an embedding layer. These embeddings are augmented with positional encodings to preserve the order of tokens, as the model itself does not inherently understand sequence order. The encoder consists of multiple identical layers (typically six). Each layer contains two main components: multi-head self-attention: This mechanism allows each token to consider other tokens in the sequence, capturing contextual relationships; and a feed-forward network: A fully connected network that processes the output of the self-attention layer by applying non-linear transformations.

The decoder section also comprises multiple layers but incorporates additional mechanisms: masked multi-head self-attention: This prevents the decoder from attending to future tokens during training, ensuring predictions are based solely on past tokens; and a cross-attention layer: This allows the decoder to focus on relevant parts of the encoder output while generating its own output sequence. The final output from the decoder passes through a linear layer and SoftMax function, converting it into a probability distribution over possible tokens, facilitating tasks such as translation or text generation. The advantages of Transformers are discussed below.

- Parallelization: Unlike RNNs which process sequences sequentially, Transformers can process all tokens concurrently, making more efficient use of modern hardware capabilities such as GPUs.
- Scalability: The architecture scales well with increasing data and model size, enabling training on massive datasets.
- Long-Range Dependencies: The self-attention mechanism allows the algorithm to capture relationships between distant tokens in a sequence, crucial for understanding context in language tasks.

Transformers have been successfully applied across diverse fields, including:

- Natural Language Processing: Used in machine translation, text summarization, sentiment analysis, and conversational agents.
- Computer Vision: Adapted for image classification and object detection tasks.
- Reinforcement Learning: Employed in models requiring sequential decision-making.

## 4.8. Generative Adversarial Networks

Generative Adversarial Networks (GANs) represent a significant advancement in machine learning, particularly in the field of generative modeling. Introduced by Ian Goodfellow et al. in 2014, GANs consist of two neural networks—a generator and a discriminator—engaged in a zero-sum game, where the success of one network comes at the expense of the other. As shown in Figure 1 the generator network is responsible for creating new data samples that mimic the training data. It takes random noise as input and transforms it into data samples, aiming to produce outputs indistinguishable from real data. The discriminator network evaluates the authenticity of the data samples it receives, distinguishing between real data from the training set and fake data generated by the generator. Its goal is to maximize its accuracy in identifying real versus generated samples.

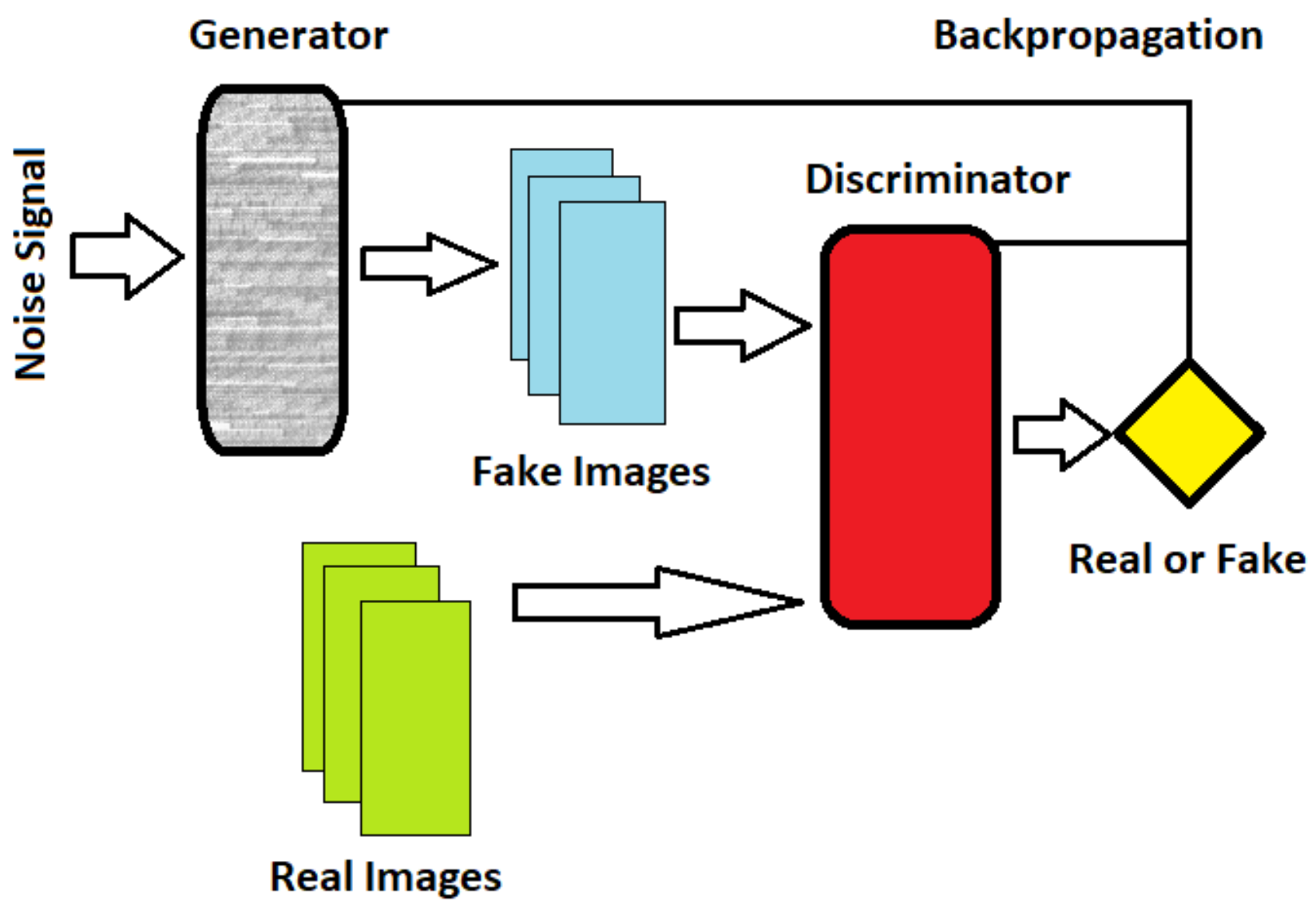

*Figure 1 Demonstration The structure and training methods of GANs*

The training process involves iteratively improving both networks: as the generator learns to produce more realistic outputs, the discriminator becomes better at detecting forgeries. This adversarial training continues until the discriminator can no longer reliably distinguish between real and generated data, ideally reaching a state where it is fooled approximately half the time. A key advantage of GANs over autoencoders lies in their broader scope for generating novel data [7]. GANs have been applied in various fields, some of which are described below. These include image generation, where GANs can create photorealistic images that do not correspond to any real person or object; image-to-image translation, used for tasks such as converting images from one style to another (e.g., summer landscapes to winter scenes); and text-to-image synthesis, where GANs can generate images based on textual descriptions, bridging natural language processing and computer vision.

### 4.9. Diffusion Models

The development of related fundamental technologies has gradually shifted from GANs to multi-step diffusion models, offering higher-quality generation capabilities. Generated content has also transitioned from single-frame images to video modeling [3]. Diffusion generative models represent a significant advancement in generative machine learning, providing a robust framework for generating high-quality data across modalities including images, audio, and text. They operate by simulating a two-step process: adding noise to a dataset and then learning to reverse this process to reconstruct the original data distribution. A diffusion model consists of three main components: a forward process, a reverse process, and a sampling method. The goal of diffusion models is to learn the diffusion process for a given dataset, such that this process can generate novel items similarly distributed to the original dataset.

The model responsible for noise removal is often referred to as its backbone. The backbone can be of various types, but they are commonly U-net or Transformer architectures. Diffusion models are predominantly used for computer vision tasks, including image denoising, inpainting, super-resolution, image generation, and video generation. These typically involve training a neural network to sequentially remove noise from progressively noisier images corrupted with Gaussian noise. The model is trained to reverse the process of adding noise to an image. After convergence during training, generation can be achieved by starting with an image composed of random noise and iteratively applying the network to denoise the image. Beyond computer vision, diffusion generative models have found applications in natural language processing (e.g., text generation and summarization), audio generation, and reinforcement learning. Diffusion models are inspired by the physical process of diffusion, where particles move from high-concentration regions to low-concentration regions. In machine learning, this concept is adapted to generate new data by systematically introducing noise into a dataset and subsequently reversing this process to recover or create new samples that resemble the original data distribution. The operation of diffusion models can be divided into two main phases:

- The forward diffusion process: This stage involves progressively adding Gaussian noise to the data. The model starts with a simple distribution and gradually complexifies it through a series of transformations, capturing intricate patterns within the data.
- The reverse diffusion process: After the data has been sufficiently corrupted with noise, the model learns to reverse this process. This stage involves training a neural network to progressively remove the noise from the data, effectively reconstructing the original distribution or generating new samples.

Diffusion models, due to their ability to generate high-quality outputs, have found applications in various fields, including: Image generation: Notable examples include DALL-E 2 and Stable Diffusion, which generate photorealistic images from textual descriptions. Audio synthesis: They are also employed in generating high-quality audio samples. Medical imaging: This technology is being explored for enhancing medical imaging techniques and diagnostics.

## 5. Deepfake Categories

Deepfake generation can be broadly categorized into four main research areas: 1) Face swapping: dedicated to implementing identity exchange between two individuals' images; 2) Face reenactment: emphasizing the transfer of source movements and gestures; 3) Talking face generation: focusing on achieving natural mouth movement synchronization with textual content in character generation; and 4) Face attribute editing: aiming to modify specific facial features of the target image. Figure 2 provides a structured overview of deepfake types and their subcategories.

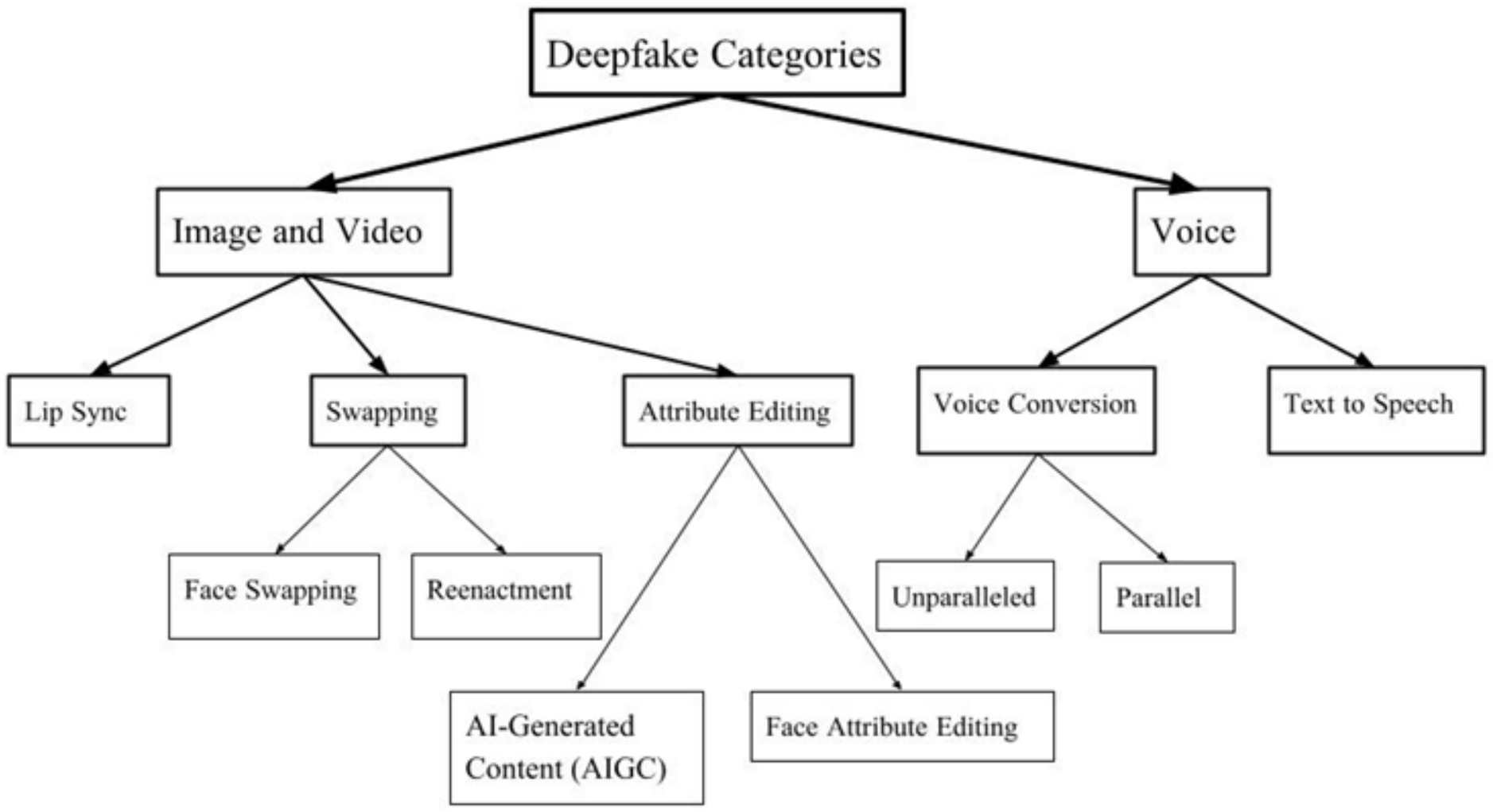

*Figure 2 Deepfake categories*

Manipulated or fabricated content can take many forms, each presenting unique challenges and posing varying risks to individuals and organizations. Different deepfakes generate manipulated content using diverse approaches. Deepfake content can be categorized as image, video, and audio deepfakes. Although image and video deepfakes might be considered synonymous, as video content is essentially a sequence of images, this paper examines manipulation of video, image, and audio content for applications such as face swapping, speaker transformation, facial expression changes, lip synchronization, and facial feature editing, each of which will be detailed separately [7 ,1].

## 5.1. Face Swapping

Face swapping technology, a prominent subset of deepfake generation, involves manipulating video content to replace one person's face with another, effectively allowing the primary subject to assume a different identity. This technique employs advanced machine learning algorithms, particularly deep learning frameworks, to achieve photorealistic results.  Generally, deepfake facial manipulations can be categorized into four main groups:

- Face generation, encompassing the creation of entirely novel facial images;
- Face attribute modification, such as altering hair color, age, gender, glasses, etc.;

- Face swapping, involving the replacement of the original person's face with another's;
- Face reenactment (sometimes referred to as expression transfer), where the original person's facial expression is transferred to a target individual.

Deepfakes can pose varying levels of risk. Among the four deepfake facial manipulations listed, face swapping and face reenactment present a significant threat to society [10]. Face swapping methods can replace the face in a reference image with the same shape and features of an input face. Deep learning methods are used to extract facial features from the input face and then transfer them to the generated face. A face in a video file may be replaced by another person while retaining the original scene content and preserving the original facial expressions [20]. Face swapping technology exemplifies the double-edged nature of deepfake advancements. It offers potential for innovative creative applications while simultaneously posing risks to social norms surrounding trust and authenticity. As this technology continues to evolve, ongoing research into detection methods and ethical guidelines will be crucial in mitigating potential harms. This section will examine face swapping methods, which can be broadly categorized into four approaches:

### 5.1.1. Traditional Graphics

Traditional face swapping remains a vital skill in graphic design and photography, offering unique advantages in customization and control. However, advancements in technology are reshaping how we approach this art form, introducing AI-powered methods that enhance creativity while streamlining workflows. Understanding both approaches allows for a more comprehensive grasp of digital image manipulation techniques, catering to diverse needs in artistic expression and professional applications. As representatives of early implementations, traditional graphic-based methods can be categorized into two approaches:

- Key point Matching and Blending: Some methods rely on matching and blending vital information to replace corresponding regions by aligning key points in the target facial areas—such as eyes, nose, and mouth—between source and target images. Subsequent steps, such as boundary blending and lighting adjustments, are then performed to produce the resulting image.

- 3D Model Construction for Face Parameterization: Methods based on 3D model construction and the introduction of a facial parameter model often involve building a facial parameter model using 3DMM technology based on a pre-collected face database. After matching the source image's facial information with the structured face model, specific modifications are made to the parameters of the facial parameter model to generate a completely new face.

### 5.1.2. Variational Autoencoders

Variational Autoencoders (VAEs) are increasingly utilized in deepfake technology, particularly for face swapping—the process of replacing a person's face in a video or image with another. VAEs play a crucial role in enhancing the realism and quality of these manipulations. Introduced in 2013, VAEs revolutionized autoencoders by altering the relationship between latent features and linear mapping within latent spaces. This involves introducing feature distributions, such as Gaussian distributions, enabling the generation of novel entities through interpolation. Therefore, VAEs are generative models that learn to encode input data into a latent space and then decode it to reconstruct the original data. This allows them to generate new data points similar to the training dataset. In face swapping, VAEs effectively capture complex facial image features, facilitating more realistic swaps. To generate deepfakes using this method, two distinct VAEs are trained—one for each face. Figure 3 illustrates the operation of a VAE; both network architectures share the same encoder but employ different decoders. These two VAEs are then used to swap face A with face B.

To swap face A with face B, face A's image is compressed using the encoder trained on face A, and face B's decoder is then used to reconstruct the image. For generating deepfake videos, this face replacement must be performed frame-by-frame [2]. Various deepfake technologies, such as that described in [4], employ this technique.

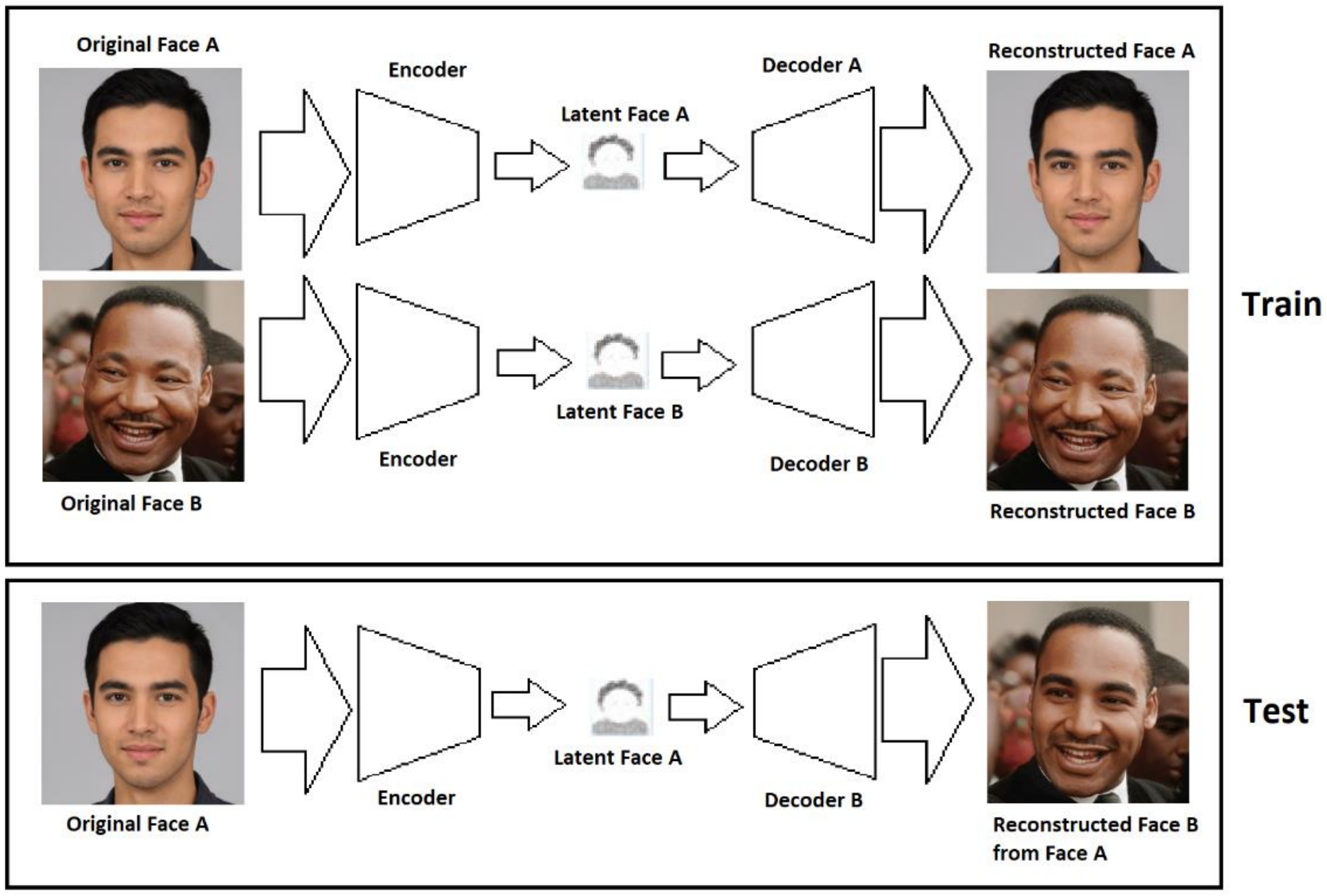

*Figure 3 The architecture and operation of Variational Autoencoders (VAEs) in training and testing phases.*

### 5.1.3. Generative Adversarial Networks

Generative Adversarial Networks (GANs) are increasingly employed in face swapping, particularly in creating deepfakes. This technology leverages the capabilities of GANs to generate realistic images by training two neural networks—a generator and a discriminator—against each other. The application of GANs to face swapping involves several key stages: face detection, alignment of pose and facial expression, and face blending. Figure 4 illustrates this face-swapping process. The goal of GAN-based methods is to obtain realistic generated images through adversarial training between the generator and discriminator network, forming the core approach to face swapping. Early GAN-based methods addressed issues related to pose and lighting consistency between source and target images. Depth Nets, a combination of GANs and 3DMM, are designed to map the source face onto any target geometry, not being limited to the target template's geometry. This makes it less susceptible to pose differences between source and target faces. However, they face challenges in generalizing the trained model to unseen faces. Significant effort has been dedicated to addressing this generalization limitation.

Recent advancements include a fine-grained editing approach termed "Edit for Swap" (E4S). This method explicitly disentangles facial shape and texture, allowing for finer control over which features are exchanged. Using component masks, this framework effectively handles occluded regions, enhancing identity preservation during the swap [21]. Regional Generative Inversion (RGI) is another technique operating in a latent space derived from StyleGAN, enabling fine-grained manipulation of facial features by replacing local shape and texture components between faces. This facilitates high-fidelity swapping that maintains realism even in complex facial poses. Another novel approach involves utilizing a ShapeEditor encoder that disentangles identity and attribute information from faces. This encoder generates encoding vectors representing these attributes, which are then used to more precisely control a StyleGAN generator. This two-stage process increases the resolution and fidelity of generated images [22].

Generalization Capability:

A deepfake network might be trained or designed to operate with only a specific set of target and source identities. Achieving an unbiased model is sometimes challenging due to correlations learned during training between target and source identities. To assess substitution or regeneration capabilities, we consider three main categories of generalization power:

- One-to-one: A model using one specific identity to drive another specific identity.
- Many-to-one: A model using any identity to drive one specific identity.
- Many-to-many: A model using any identity to drive any other identity [23].

### 5.1.4. Diffusion Models

Diffusion models have emerged as powerful tools in generative AI, particularly for creating deepfakes, including face-swapping applications. These models operate by progressively corrupting training data with noise and then learning to reconstruct the original data through a reverse process. This capability allows them to generate highly realistic images from text prompts or existing images. Recent advancements have specifically adapted diffusion models for deepfake generation, enhancing their ability to create believable visual content consistent with context. For instance, modifications to established models like Stable Diffusion enable the generation of high-quality deepfakes [24].

In face-swapping applications, diffusion models employ a combination of techniques to ensure high quality and realism. A typical framework may incorporate components such as image feature encoding, conditional generation, and inpainting for seamless replacement of facial features while preserving surrounding texture. Using advanced face-guided optimization methods, these models can maintain identity consistency in swapped faces. Some techniques further refine the output by training on specific features associated with the target identity, resulting in a more coherent and realistic final image. The integration of these methodologies highlights the sophisticated nature of modern deepfake creation, where even subtle details can be manipulated to achieve a high degree of realism. The proliferation of generative models for deepfake creation presents significant challenges, particularly regarding detection. The increasing realism and diversity of content generated by these models complicate the identification of manipulated media, forcing existing detection systems to adapt effectively. Research indicates that traditional detection methods often struggle with the intricate nature of deepfakes produced by diffusion models, which can differ significantly from previously encountered manipulations. In response, efforts are underway to enhance detection capabilities through improved training data diversity and innovative strategies that account for the unique characteristics of diffusion-generated content. This ongoing arms race between generation and detection highlights the critical need for robust solutions to mitigate the risks associated with deepfake technology.

**Table 1- Detailed description of the tools reviewed for creating image and video deepfakes.**

| Toolkit | Release Date | Repository | Doc | Performance | Written in | Stars | Collection |
|---|---|---|---|---|---|---|---|
| **roop-unleashed** | 2023 | GitHub | DOC | ★★★★★ | Python | 1.6 | faceswap |
| **LivePortrait** | 2024 | GitHub | LivePortrait (hf demo) | ★★★★★ | Python | 11.7 | reenactment, video-editing, |
| **AniTalker** | 2024 | GitHub | hf demo | ★★★★★ | Python | 1.3 | Audio-Driven Portrait Animations, lip-sync |
| **Wav2Lip** | 2021 | GitHub | | ★★★★☆ | Python | 9.9 | lip-sync |
| **Roop** | Jun 25, 2023 | GitHub | README file | ★★★★☆ | Python | 25 | faceswap |
| **Deep-Live-Cam** | 2023 | GitHub | | ★★★☆☆ | Python | 39 | real time deepfake, video-deepfake |
| **V-Express** | 2024 | GitHub | project page | ★★★☆☆ | Python | 2.1 | Audio-Driven Portrait Animations |
| **DreamTalk** | 2024 | GitHub | | ★★★☆☆ | Python | 1.5 | Audio-Driven Portrait Animations |
| **EchoMimic** | 2024 | GitHub | hf demo | ★★★☆☆ | Python | 2 | Audio-Driven Portrait Animations |
| **AniPortrait** | 2024 | GitHub | hf demo | ★★★☆☆ | Python | 4.4 | Audio-Driven Portrait Animations |
| **SadTalker** | 2023 | GitHub | hf demo | ★★★☆☆ | Python | 11.4 | image-animation, talking-head |
| **video-retalking** | 2023 | GitHub | | ★★★☆☆ | Python | 6.2 | Talking Head Video Editing, lip sync |
| **facefusion** | 20 Aug 2023 | GitHub | DOC | ★★☆☆☆ | Python | 8.6 | faceswap |
| **hallo** | 2024 | GitHub | tts demo on hf | ★★☆☆☆ | Python | 8 | Audio-Driven Visual Synthesis |
| **StyleHEAT** | 2022 | GitHub | | ★★☆☆☆ | Python | 0.6 | editable talking face generation |
| **HyperReenact** | 2023 | GitHub | | ★★☆☆☆ | Python | | Reenactment |
| **Thin-Plate-Spline-Motion-Model** | 2022 | GitHub | hf demo | ★☆☆☆☆ | Python | 3.4 | face-reenactment |
| **styletalk** | 2023 | GitHub | | ★☆☆☆☆ | Python | 0.5 | Talking Head Generation |
| **headswap-machine** | 2021 | GitHub | | ★☆☆☆☆ | Python | | headswap |
| **SimSwap** | June 11 2021 | GitHub | DOC | ☆☆☆☆☆ | Python | 3.6 | Faceswap, image manipulation |
| **ReliableSwap** | 2023 | GitHub | DOC | ☆☆☆☆☆ | Python | | faceswap |
| **Wav2Lip-GFPGAN** | 2021 | GitHub | | ☆☆☆☆☆ | Python | | lip-sync, wav2lip |
| **inswapper** | 2023 | GitHub | | ☆☆☆☆☆ | Python | | faceswap |
| **faceswap-GAN** | 2018 | GitHub | README file | ☆☆☆☆☆ | Python/Jupyter Notebook | 3.3 | Faceswap, image manipulation |
| **fewshot-face-translation-GAN** | 2019 | GitHub | README file | ☆☆☆☆☆ | Python/Jupyter Notebook | 0.7 | Faceswap, image translation |
| **dfaker** | 2017 | GitHub | README file | ☆☆☆☆☆ | Python | 0.4 | faceswap |
| **speech-driven-animation** | 2020 | GitHub | | ☆☆☆☆☆ | Python | 0.9 | Speech-Driven Animation |
| **SadTalker-Video-Lip-Sync** | 2023 | GitHub | | ☆☆☆☆☆ | Python | 1.8 | lip-sync |
| **Moore-AnimateAnyone** | 2024 | GitHub | hf demo | ☆☆☆☆☆ | Python | 3 | Face Reenactment |
| **facechain** | 2024 | GitHub | | ☆☆☆☆☆ | Python | 8.8 | portrait generation |

## 5.2. Reenactment

Facial expression transfer, or face reenactment, is a sophisticated technique in the realm of deepfakes that refers to the process of transferring facial expressions and movements from one person (the driver) onto the face of another (the driven) while preserving the original identity [10 ,3]. This method allows for the seamless and realistic portrayal of emotions and expressions, distinguishing it from other deepfake techniques such as face swapping, where one person's face is entirely replaced with another's. This technology utilizes advanced machine learning algorithms, specifically Generative Adversarial Networks (GANs) and Variational Autoencoders (VAEs), to achieve high fidelity in the reenacted facial features, resulting in a convincing illusion of expression transfer.

The beneficial applications of face reenactment are diverse, spanning entertainment, advertising, and even political contexts. For example, it can be used in film production to create expressive characters or dub foreign films with accurate lip synchronization. Furthermore, it has implications for virtual reality and gaming, where realistic avatars can mimic human emotions in real-time. However, these advancements also raise significant ethical concerns regarding misinformation and identity theft, as the technology may be misused to create misleading or harmful content. The ability to convincingly manipulate visual media presents challenges to authenticity in digital communication. Unlike face swapping techniques, facial expression transfer techniques are rarely considered in readily available datasets. Reference [9] is a significant and commonly used resource for facial expression transfer, employed in numerous works. In this work, facial landmarks are first extracted using 3D detectors to provide representative images for the source and target faces. Then, low-dimensional representations of parameters such as pose and expression from the source, and style information from the target video, are extracted using an encoder network. Research on face reenactment continues to evolve, focusing on improving the quality and efficiency of these techniques while addressing their ethical implications. Recent studies emphasize the importance of identity preservation throughout the reenactment process, as unwanted leakage of identity information from the driver video can compromise the integrity of the output.

### 5.3. Lip sync

Deepfake lip synchronization represents a sophisticated application of artificial intelligence in the manipulation of digital media. These deepfakes involve altering a person's lip movements to match a new or modified audio track, creating the illusion that they are speaking words they did not originally utter. This technology is particularly concerning because it focuses on a localized area—the mouth—making inconsistencies harder to detect compared to full-face manipulations. Recent advancements have highlighted the challenges associated with identifying this type of deepfake, as they can often appear more convincing than traditional face-swapping techniques due to the localized nature and subtleties involved in lip synchronization. Deepfake audio tools possess the capability to generate highly realistic lip-synchronized videos where audio features are manipulated and aligned with video frames [7].

Generally, lip synchronization can be considered a temporal mapping aiming to generate a talking video where the target image's character speaks based on an arbitrary motion source, such as text, audio, video, or a combination of multimedia sources. The lip movements, facial expressions, emotions, and speech content of the generated video's character match the target information [3]. Recent research has focused on developing methods to identify these deceptive videos by analyzing inconsistencies between audio and visual components. A notable approach is the LIP-INCONS model, which detects temporal inconsistencies in mouth movements across frames [25]. This model captures both local and global mouth features to assess whether lip movements align with the spoken words. By examining these discrepancies, researchers have significantly improved detection accuracy, outperforming existing techniques on various benchmark datasets. The effectiveness of such methods is crucial given the increasing prevalence of lip-synchronized deepfakes across various media.

### 5.4. Face Attribute Editing

Facial attribute manipulation in the context of deepfakes refers to the process of altering specific facial attributes using advanced deep learning techniques, particularly Generative Adversarial Networks (GANs). This manipulation can encompass changes in age, gender, ethnicity, and other characteristics such as skin texture, hair color, and even emotional expressions. Such capabilities have become increasingly accessible due to the proliferation of user-friendly applications like FaceApp and similar tools, enabling even non-experts to create realistic alterations in facial images and videos. Existing methods encompass both single-attribute and comprehensive editing: the former focuses on training a model for a single attribute, while the latter integrates multiple feature editing tasks simultaneously—the primary focus of this review [3].

The technological underpinnings of facial feature manipulation involve sophisticated algorithms capable of blending or altering facial features while maintaining a high degree of realism. For example, feature manipulation techniques often utilize conditional GANs that enable targeted alterations based on user-defined attributes. This process involves an encoder that captures the principal features of the face and a GAN that generates the modified output. The result is a highly convincing image or video where specific features have been altered without losing the overall coherence of the subject's appearance. This capability raises concerns regarding the integrity of biometric systems and automated recognition technologies, which may struggle to accurately identify manipulated faces. Automated facial recognition systems exhibit significant vulnerabilities when confronted with manipulated images. Studies have shown that error rates can dramatically increase under various forms of manipulation. For instance, manipulations such as digital reshaping or beautification can, in some cases, lead to identification failures of up to 95%. Consequently, ongoing research focuses on developing robust detection methods capable of effectively identifying manipulated content while also addressing ethical concerns surrounding privacy and consent in the digital age. The dual nature of these technologies—as both creative tools and potential instruments of deception—highlights the need for comprehensive understanding and responsible usage.

**Table 2- Detailed description of the tools reviewed for Face Attribute Editing.**

| Toolkit | Release Date | Repository | Doc | Performance | Written in | Stars | Collection |
|---|---|---|---|---|---|---|---|
| **Inpaint-Anything** | 2023 | GitHub | | ★★★★★ | Python/Jupyter Notebook | 4.7 | Inpainting |
| **lama-cleaner** | 2021 | GitHub | | ★★★★★ | Python | 14 | Inpainting |
| **CodeFormer** | Aug 10, 2022 | GitHub | DOC | ★★★★★ | Python | 10.9 | face-restoration, Face Inpainting |
| **latent-diffusion** | 2022 | GitHub | | ★★★★☆ | Python | 9.4 | text2image - inpainting |
| **ProPainter** | 2020 | GitHub | DOC | ★★☆☆☆ | Python | 4.1 | Video Inpainting |
| **RePaint** | 2022 | GitHub | | ★☆☆☆☆ | Python | 1.6 | inpainting |
| **lama** | 2023 | GitHub | | ★☆☆☆☆ | Python/Jupyter Notebook | 6.7 | Inpainting |

## 5.5. Voice Conversion

Artificial speech generation can be broadly categorized into two types: text-to-speech (TTS) and voice conversion (VC). In TTS, the AI model receives text data as input and generates the corresponding synthetic speech. VC, however, allows users to transform one person's voice into another's. In TTS, all generative audio models, at their most basic level, utilize an encoder-decoder-like concept. Thus, in the first stage, the encoder must encode text features into acoustic and prosodic features. Then, in the second stage, it must decode these features and generate the speech waveform. Consequently, GAN-based approaches align well with the discussion of audio generation. Modified versions of CycleGAN and StarGAN have also been proposed for synthetic speech generation tasks [7]. Voice conversion (VC) is a technology enabling the transformation of one speaker's voice into another's while preserving the original content. Two main approaches exist for achieving voice conversion: parallel and non-parallel training, each possessing distinct characteristics, advantages, and challenges. In parallel voice conversion, both the source and target speakers read the same sentences during the data collection phase. This relies on a parallel corpus, meaning that for every utterance produced by the source speaker, there is a corresponding utterance by the target speaker. This alignment facilitates accurate feature extraction and mapping between voices, leading to higher-quality synthesized speech. However, collecting sufficient parallel data can be challenging, as both speakers must produce identical speech content, often impractical in real-world scenarios [26].

Conversely, non-parallel voice conversion does not require the source and target speakers to read identical sentences. Instead, it utilizes a diverse collection of utterances from both speakers that may not be directly aligned. This approach is particularly beneficial when parallel data is scarce or unavailable. Non-parallel methods allow for greater flexibility in training data collection, as it does not necessitate identical utterances from both speakers. This can be applied in various contexts where parallel data acquisition is infeasible, such as multilingual scenarios or those with limited speech samples. However, the lack of direct alignment can lead to reduced quality in the synthesized speech, as the model may struggle to accurately map features between mismatched utterances. Non-parallel methods often require more sophisticated algorithms and techniques to bridge the gap between different speaking styles and patterns [27]. Open-source resources available on platforms such as GitHub were used to identify suitable and reliable sources, as well as programming tools and code for implementing the methods. Table 2 presents a list of all the relevant collected tools along with supplementary information. Several key metrics are commonly used to evaluate voice conversion models. These include:

- Multilingual capability: The ability to convert voices between different languages.
- Low-resource learning: The ability to train effectively with limited source data.

- Quality: The assessment of output audio quality, primarily qualitative in nature.
- Speed: Generally, less complex models exhibit faster processing speeds.
- Extensibility: The ease with which the software can be further developed.
- Up-to-dateness: The model's use of the latest advancements in the field.

The ordering in Table 1 reflects the results of our evaluations, comparing model outputs and considering the aforementioned metrics. The model presented by tool number one in Table 1 is based on a k-nearest neighbor algorithm and the extraction of fundamental acoustic signal features [28]. This method first extracts the reference and target audio signals using a pre-trained reference tool [29]. Then, each frame of the reference audio feature vector is replaced with its nearest neighbor in the target audio feature vectors. Finally, a pre-trained generative audio model [30] produces the target audio signal from the modified feature vectors. A key advantage is its text-independence. Some voice conversion models utilize text from the source audio transcript in the bottleneck layer, but this adds computational overhead and eliminates useful information from the reference audio signal, such as speaker intonation, pitch, identity, and timing. Recent work confirms that higher layers of neural networks exhibit poorer representation of this information. Based on these observations, it's determined that using a highly correlated layer (layer 6 in the reference model [30]) is crucial for good speaker similarity and preserving prosodic information from the source audio signal. Therefore, features extracted from layer 6 were used for all experiments, producing one vector for every 20 milliseconds of 16 kHz audio.

Regarding training data volume, increasing the data size enhances the system's ability to detect and transfer the target audio's acoustic features. Longer target audio files generally improve output quality, although this improvement is limited; for files exceeding 5 minutes, the difference becomes negligible. Conversely, training data shorter than 30 seconds significantly reduces model performance [28]. Using this method, the audio generator produces higher-quality audio. This improvement stems from the model's use of the mean of each frame's components during testing, creating a form of alignment between training and testing. Due to its non-parametric nature, the model demonstrates relatively good accuracy in converting different languages, whispers, and even non-human sounds.

**Table 3- Detailed description of the tools reviewed for Voice Conversion.**

| Toolkit | Release Date | Repository | Doc | Performance | Written in | Stars | Collection |
|---|---|---|---|---|---|---|---|
| **seed-vc** | 2024 | GitHub | | ★★★★★ | Python | | voice conversion - singing voice conversion |
| **knn-vc** | 2023 | GitHub | DOC | ★★★★☆ | Python | 0.4 | Voice Cloning - Voice Conversion |
| **HierSpeechpp** | 2023 | GitHub | | ★★★★☆ | Python | 1.2 | text-to-speech (TTS), voice conversion (VC) |
| **metavoice-src** | 2024 | GitHub | | ★★★★☆ | Python | 1.2 | Zero-shot cloning |
| **StyleTTS2** | 2023 | GitHub | DOC | ★★★☆☆ | Python | 4.6 | Text-to-Speech |
| **bark-gui** | 2023 | GitHub | | ★★★☆☆ | Python | | generative-audio - TTS |
| **Bark-Voice-Cloning** | 2023 | GitHub | | ★★★☆☆ | Python | 1.8 | Voice Conversion |
| **FreeVC** | 2022 | GitHub | DOC | ★★★☆☆ | Python | | Voice Conversion |
| **TriAAN-VC** | 2023 | GitHub | | ★★☆☆☆ | Python | | Voice Conversion |
| **Real-Time-Voice-Cloning** | 2019 | GitHub | DOC | ★★☆☆☆ | Python | 52 | voice cloning |
| **so-vits-svc-fork** | 2023 | GitHub | | ★★☆☆☆ | Python | 7.9 | Voice Conversion |
| **coqui-ai/TTS** | Mar 9, 2021 | GitHub | DOC | ★★☆☆☆ | Python | 33 | Voice Cloning - Voice Conversion |
| **ppg-vc** | 2020 | GitHub | | ★★☆☆☆ | Python | | Voice Conversion |
| **QuickVC** | 2023 | GitHub | | ★☆☆☆☆ | Python | | Voice Conversion |
| **ConsistencyVC-voive-conversion** | 2023 | GitHub | | ★☆☆☆☆ | Python | | cross-lingual voice conversion and expressive voice conversion |
| **lvc-vc** | 2023 | GitHub | | ☆☆☆☆☆ | Python | | Voice Conversion |
| **StyleTTS-VC** | 2023 | GitHub | | ☆☆☆☆☆ | Python | | Voice Conversion |
| **YourTTS** | Nov 19, 2021 | GitHub | DOC | ☆☆☆☆☆ | Python | | Voice Conversion |

# 6. Deepfake Detection

As deepfake technology continues to evolve, so do the challenges associated with identifying these manipulations. The emergence of deepfake technology presents significant ethical challenges, particularly concerning misinformation, identity theft, and privacy violations. As the technology becomes more accessible, it is crucial for researchers and developers to implement robust detection mechanisms alongside generation techniques to mitigate potential misuse. Deepfake detection aims to identify anomalies, manipulations, or forged regions in images or videos using anomaly detection probabilities, holding high research and practical value in information security and multimedia forensics [3].

Recent advancements in deepfake detection utilize deep learning architectures, particularly Convolutional Neural Networks (CNNs) and Long Short-Term Memory networks (LSTMs). These models analyze both spatial and temporal features in video frames to identify inconsistencies indicative of manipulation. For example, techniques such as optical flow analysis and temporal pattern recognition are employed to detect anomalies in facial movements and background consistency across frames. Research indicates that hybrid approaches combining CNNs with LSTMs can achieve detection accuracy exceeding 95% on benchmark datasets like FaceForensics++ and Celeb-DF, demonstrating effectiveness in identifying deepfake content even in challenging scenarios. Deepfake detection methods can be broadly categorized into two types: fake face detection and AI-Generated Content (AIGC) detection, the latter encompassing a much wider scope. Fake face manipulation can be localized, affecting only the facial region, while AIGC artifacts can be global, encompassing the entire synthesized image content. Consequently, most detection methods are applicable to only one of these problem types [5]. The following sections briefly introduce each of these methods.

### 6.1. Fake Face Detection

Initial deepfake detection techniques primarily relied on traditional digital image processing methods. These approaches utilized inherent video characteristics such as inconsistencies in lip synchronization and gaze, depth of field, color components, and head poses. However, with the advent of deep learning, the focus has shifted towards learning-based detectors due to their superior feature extraction capabilities [5].

### 6.2. AI-Generated Content (AIGC) Detection

This encompasses a broader range of techniques aimed at identifying synthetic content that may not be limited to facial manipulations. Unlike fake face detection, which primarily focuses on discrepancies within facial features, AIGC detection targets a wider spectrum of artifacts generated by various AI models, including text, images, and videos. AIGC detection methods can be categorized into several approaches, some of which are detailed below:

- Feature-Based Methods: These techniques analyze inherent content features. This might involve examining statistical properties of the image or video, such as pixel distribution, noise patterns, and compression artifacts. For instance, Error Level Analysis (ELA) is a common method that highlights pixel intensity differences to detect alterations in an image.

- Machine Learning Approaches: Traditional machine learning techniques utilize handcrafted features for classification. However, these methods often face challenges in generalizing to unseen data and are sensitive to variations in input quality. Recent advancements have integrated more sophisticated algorithms such as Support Vector Machines (SVMs) and K-Nearest Neighbors (KNN) to improve classification accuracy.

- Deep Learning Techniques: The most promising advancements in AIGC detection stem from deep learning methods. Convolutional Neural Networks (CNNs) and Recurrent Neural Networks (RNNs) are employed to automatically extract complex features from data without manual intervention. For example, CNNs can capture spatial hierarchies in images, while RNNs can analyze temporal sequences in videos. Some models utilize Long Short-Term Memory (LSTM) networks to identify frame inconsistencies in video content.

- Temporal Analysis: This approach focuses on the temporal coherence of video frames. By analyzing frame sequences instead of individual images, detection systems can identify unnatural movements or inconsistencies that may indicate manipulation. Recent studies have shown that combining spatial and temporal feature analysis significantly improves detection rates.

- Hybrid Approaches: Some advanced systems combine multiple techniques to improve robustness against diverse types of AIGC. These hybrid models may integrate feature extraction methods with machine learning classifiers or utilize ensemble learning techniques that leverage the strengths of different algorithms.

Table 3 lists the key tools employed in this research. Most existing deepfake detectors can be broadly categorized into three types: preliminary, spatial, and frequency-based. The first type, preliminary detectors, utilize CNNs to perform binary classification, distinguishing fake content from authentic data. Several CNN-based binary classifiers have been proposed, such as MesoNet and Xception. The second type, spatial detectors, place greater emphasis on representations such as the location of the forged region, discriminative learning, image reconstruction, inpainting, image blending, etc. Finally, the third type, frequency-based detectors, address this limitation by focusing on the frequency domain for forgery detection [5].

**Table 4- Detailed description of the tools reviewed for Deepfake Detection.**

| Toolkit | Release Date | Repository | Doc | Performance | Written in | Collection |
|---|---|---|---|---|---|---|
| **videofact-wacv-2024** | 2024 | GitHub | | ★★★★★ | Python | Video Forgeries Detection |
| **ImageForensicsOSN** | 2022 | GitHub | | ★★★★★ | Python | Image Forgeries Detection |
| **deepfake-image-detection** | 2024 | | **(hf demo)** | ★★★★☆ | Python | Image Forgeries Detection |
| **NPR-DeepfakeDetection** | 2024 | GitHub | **(hf demo)** | ★★★★☆ | Python | Image Forgeries Detection |
| **Fake_Face_Detection** | 2024 | | **(hf demo)** | ★★★☆☆ | Python | Fake Face Detection |
| **CNNDetection** | 2020 | GitHub | | ★★★☆☆ | Python | Image Forgeries Detection |
| **FOCAL** | 2023 | GitHub | | ★★★☆☆ | Python | Image Forgery Detection and Localization |
| **SOFTX-D** | 2023 | GitHub | | ★★★☆☆ | Python | Video and Image Forgeries Detection |
| **SeqDeepFake** | 2022 | GitHub | | ★★★☆☆ | Python | Video Forgeries Detection |
| **audio-visual-forensics** | 2023 | GitHub | DOC | ★★☆☆☆ | Python | Video Forgeries Detection |
| **FACTOR** | 2023 | GitHub | | ★☆☆☆☆ | Python | Audio-Visual and Face-Forgery Detection |
| **HiFi_IFDL** | 2023 | GitHub | | ★☆☆☆☆ | Python | Image Forgery Detection |
| **FTCN** | 2021 | GitHub | | ★☆☆☆☆ | Python | Video Face Forgery Detection |
| **RealForensics** | 2022 | GitHub | | ☆☆☆☆☆ | Python | Face Forgeries Detection |
| **ICT_DeepFake** | 2022 | GitHub | | ☆☆☆☆☆ | Python | Face Forgeries Detection |
| **Face-Liveness-Detection-SDK** | 2023 | HF | DOC | ☆☆☆☆☆ | Python | Face Forgeries Detection |
| **VideoDeepFakeDetection** | 2024 | GitHub | | ☆☆☆☆☆ | Python | Video Forgeries Detection |
| **MVSS-Net** | 2021 | GitHub | | ☆☆☆☆☆ | Python | Image Manipulation Detection |

## 7. Conclusions

This paper examines the generation and detection of deepfake audio and visual content, focusing on modern deep learning technologies and related tools. The emergence of deepfakes as a novel technology presents unprecedented challenges in areas such as privacy, information security, and media credibility. However, deepfake technology also offers potential benefits, capable of enhancing various digital industries and providing innovative applications in entertainment, education, and digital communication. This research reveals the dual nature of deepfakes: while offering new possibilities in content creation, they also pose significant threats to society, national security, and individual rights. This dual nature necessitates a comprehensive approach to understanding its implications and developing effective countermeasures. Therefore, the detection and identification of deepfakes are crucial. The ease of access to deepfake creation tools, coupled with the increasing sophistication of generation techniques, underscores the urgent need for robust and readily deployable detection methods. Current detection strategies, while improving, still face challenges in identifying increasingly realistic deepfakes, highlighting the need for further research into more robust and generalizable detection algorithms. The analyses presented in this paper, along with the review of various techniques, aim to familiarize readers with the challenges and available tools in this field, enabling them to understand both the potential positive and negative impacts of deepfake technology. This paper first discusses current deep learning methods and tools widely used to create deepfake images and videos. We then examine various types of deepfakes, including face swapping, voice conversion, facial expression manipulation, lip synchronization, and facial feature editing. Furthermore, we provide a comprehensive overview of diverse technologies and their application in deepfake detection. Based on the current state-of-the-art and prevalent trends, future research should focus on several key areas:

- Advanced Detection Algorithms: The development of more sophisticated algorithms capable of identifying deepfake content with higher accuracy is crucial. This involves leveraging advancements in artificial intelligence and machine learning to improve detection rates across various media formats.
- Exploration of novel features and techniques for deepfake detection beyond current methods, potentially incorporating physiological signals or subtle artifacts.
- Development of user-friendly and accessible deepfake detection tools for the general public.
- Public Awareness and Education: Raising public awareness about deepfakes is paramount. Educational initiatives should inform individuals about the existence and implications of deepfakes and equip them with critical thinking skills to discern authentic content from manipulated material.
- Interdisciplinary Collaboration: Future endeavors should involve interdisciplinary collaboration—combining insights from computer science, psychology, law, and ethics—to holistically address the multifaceted challenges arising from deepfakes.
- Longitudinal Studies: Conducting longitudinal studies on the societal impacts of deepfakes will provide valuable insights into their evolving role in media consumption and public discourse.

By addressing these areas, researchers can contribute to a safer digital environment that balances technological innovation with ethical considerations, ultimately protecting individual rights and societal integrity against the backdrop of rapidly advancing capabilities in deepfake technology.